%Paper: hep-ph/9310273
%From: zhangx@umdhep.umd.edu
%Date: Tue, 12 Oct 1993 13:34:38 EDT

\font\titlefont = cmr10 scaled \magstep2
\magnification=\magstep1
\vsize=20truecm
\voffset=1.75truecm
\hsize=14truecm
\hoffset=1.75truecm
\baselineskip=20pt

\settabs 18 \columns

\def\b{\bigskip}
\def\bb{\bigskip\bigskip}
\def\bbb{\bigskip\bigskip\bigskip}

\def\ce{\centerline}

\def\no{\noindent}

%$$\eqalign{
% put in lines of equations here, each ending in \cr
%}$$

%$$\eqalign{
%put in equations here ending each line with \cr
%} \eqno (1)$$
%the above will put the one between the two lines of equations and set it
%off to the right

% END BEGINNING FORMATS
% BEGIN HEADER

 \rightline{ UMDHEP 94-42}
\rightline{ October 1993}
\b
\ce{\titlefont{
Effects of $ e^+ e^- \nu_e$ Decays of
 }}
\ce{\titlefont{ Tau Neutrinos Near A Supernova}}

\bb
\ce{\bf{R.N. Mohapatra\footnote\dag{\rm{ Work supported by a grant from the
National
          Science Foundation}},~~
S. Nussinov\footnote{*}{\rm{Permanent address: Department of Physics and
Astronomy, Tel Aviv University, Tel Aviv, Israel.}}
  and X. Zhang{$\dag$}\footnote{**}{{\rm {
  Address after November 1, 1993: Department of
Physics and Astronomy, Iowa State University, Ames, Iowa 50011}}}
   }}

\ce{\it{ Department of Physics and Astronomy}}
\ce{\it{University of Maryland}}
\ce{\it{ College Park, MD 20742 }}
\b
\ce{\bf Abstract}

We revisit the constraints implied by
SN 1987A observations on the decay rate of a multi-MeV $\nu_\tau$
decaying into the visible channel $\nu_\tau \rightarrow
          e^+ e^- \nu_e$, if its lifetime is more than 10 {\it sec}.
We discuss its implication for the minimal left-right symmetric model with
see-saw mechanism for neutrino masses. We also
speculate on the possible formation of a ``giant Capacitor"
in intergalactic space due to the decay of ``neutronization"
$\nu_\tau$'s and spin allignment possibility in the supernova.

\no
 \filbreak
\no {\bf I. Introduction}~~~~~

One of the most important
questions in particle physics today is whether
the neutrinos have mass[1]. A non-zero neutrino mass would indicate rich
new physics beyond the standard model. Several attractive scenarios for
such physics already exist, predicting neutrino masses much smaller than
those of the corresponding
charged fermions via the see-saw mechanism[2]. In these theories,
 the neutrino mass provides an estimate of the scale ($V_{BL}$) above which
global or local B-L symmetry will manifest. In the usual see-saw
models with local B-L symmetry[3],
the neutrino masses scale quadratically with the
charged lepton
(or quark)
masses of the corresponding generation. If we entertain
the possibility that the scale of local B-L symmetry is in the TeV range,
( a possibility completely consistent with all known data ), then
quadratic mass formulae mentioned above
predict
the tau neutrino mass to be in the MeV range (
and the $\nu_\mu$ and
$\nu_e$ in the KeV and eV ranges respectively).
Accelerator data[4] provides an
 upper limit of 31 MeV on $m_{\nu_\tau}$.
A tau neutrino in the MeV range has the unique property that it can decay to
visible channels such as
$\nu_\tau \rightarrow \nu_e ~e^+ ~ e^-$. This decay channel
has interesting implications in many situations.
 In this paper, we
study one of them, which arises near SN 1987A, and
discuss the constraints on this
decay mode of $\nu_\tau$
 using SN 1987A observation.

It is well-known that
 if the mass of the tau neutrino $\nu_\tau$
lies in the six decade range:

$${
30~ {\rm eV} \leq m_{\nu_\tau} \leq 31~ {\rm MeV}    ~~~~,}
\eqno(1)$$
\no
then it should decay in order to be consistent with
the standard big bang cosmology.
  This decay
must be fast enough[5] ( for $m_{\nu_\tau} \leq {\rm few ~~MeV}$ )
$${
\tau_{\nu_\tau}  \leq 5.4 \times 10^8 {( { {\rm MeV} \over m_{\nu_\tau} })}^2
{}~{\it sec} ~~~, }
\eqno(2)$$
in order that the relic $\nu_\tau$ decay
products, even if massless and unobservable, red shift sufficiently
to have
their contribution to the energy density of the universe remain
below the critical density {\it i.e}
 $\rho ( \nu_\tau~~ {\rm decay~~ products} ) \leq
\rho_{crit}$. Demanding in addition that the
$\nu_\tau$ decay will not disturb structure (Galaxy) formation leads to the
more model dependent ( hence less reliable ) bound[6],
$${
\tau_{\nu_\tau} \leq 10^{4} ~{\it sec}~ {( {1 ~{\rm MeV} \over m_{\nu_\tau}}
)}^2   ~~~~.}\eqno(3)$$

The possible decay modes
of $\nu_\tau$ in the extensions of standard model are[1]:
$${
\nu_\tau \rightarrow \nu_i~~ \nu_j~~ \nu_k , ~~~~( i, j, k = e, \mu ) ~,}
\eqno(4)$$
\no the three neutrino mode, the radiative mode
$${
 \nu_\tau \rightarrow \nu_i ~+~ \gamma ~~~,}
\eqno(5)$$
\no and if $m_{\nu_\tau} \geq 2~ m_e ~ \simeq 1 $MeV, the three body
charged mode
\footnote{[F.1]}{
In certain models with spontaneous breaking of global B-L[7], we have
massless singlet
majorons J and $\nu_\tau \rightarrow \nu_i~ + ~J$ is possible.
In some versions of these models, $\nu_\tau$ lifetime could be made very short
( {\it i.e.} $\leq 1~ {\it sec}$ ).  We will not consider such models here;
 instead we discuss models, where $\nu_\tau$ lifetime is bigger than 10
{\it sec}, which,
for example, may happen in the minimal singlet Majoron model ( see later ).},

$${
\nu_\tau \rightarrow \nu_e~~ e^- ~~e^+ ~~~~~.}
\eqno(6)$$
There are no standard model tree diagrams contributing to
the decay mode in eq.(4) into the neutrinos.
In the left-right symmetric model, exchange of Higgs bosons lead to
this decay mode at the tree level
with a rate given by
$${
\Gamma ( \nu_\tau \rightarrow \nu_i \nu_j \nu_k ) =
{\tilde G_F}^2 m_{\nu_\tau}^5
            / 192 \pi^3  ~~~~~~, }\eqno(7)$$
\no where ${\tilde G_F} \simeq {\sqrt 2}
f_{\tau i}f_{j k} /4~ m_{\Delta^0}^2$. Making the reasonable
assumption that
$${
    {\tilde G_F} \leq G_F ~~~~~~~,}\eqno(8)$$
\no with $G_F = 10^{-5} / m_N^2$ the ordinary Fermi constant,
we find that if the rate of the three neutrinos decay
mode satisfies the bounds (2)
and (3), we need to have respectively[9, 10]
$${
m_{\nu_\tau} \geq 0.035 ~~{\rm MeV~~and}~~
m_{\nu_\tau} \geq 0.1 ~~{\rm MeV}
{}~~~~.}\eqno(9)$$

Turning to the radiative decay, it involves the transition of
magnetic moments
$\mu_{\nu_e \nu_\tau}.  ~~ {\rm Defining} ~~\mu_{\nu_e \nu_\tau} \equiv
\kappa_{e \tau} e \hbar /2 m_e c ,$ we know that
they are bounded by
$\kappa_{i \tau} \leq 10^{-9}$ from
 direct $\nu_i ~e^-$ scattering experiments[11].
Since the radiative decay rate are $\Gamma( \nu_\tau \rightarrow \nu_i \gamma)
 \simeq { \kappa_{ \tau i}^2 \alpha \over { 16 \pi^2} }
{ m_{\nu_\tau}^3 \over m_e^2}
$,
one finds that eq.(2) is satisfied for all allowed values of $m_{\nu_\tau}$
(down to the eV range) whereas eq.(3) requires $m_{\nu_\tau} \geq .3$ KeV.
However, once the $\nu_\tau$ mass is in the range of few KeV's, the
astrophysical constraints on $\mu_{\nu_\tau \nu_e}$, which are much more
stringent ( {\it i.e.} $\kappa_{\tau e} \leq 10^{-12}$ ) become
operative[12]. Thus, the effective lower bound on $m_{\nu_\tau}$
from radiative decays is in the few KeV range.
One must of course realize that from a theoretical standpoint
$\kappa_{\tau i} \simeq 10^{-9}$ may not be so easy to obtain as
$m_{\nu_\tau}$ becomes lighter than few KeV's.
In any case, once
 $m_{\nu_\tau} \geq $ MeV,  the
$\nu_\tau \rightarrow e^+ e^- \nu_e$ decay mode
arises and we focus on this from now on.
 Such a decay could
result from a standard model W exchange diagram
provided one adds the right-handed neutrinos to the
model and allows for neutrino mixing:
 $${
\Gamma_{SM}^{( \nu_\tau \rightarrow \nu_e e^+ e^- )} \simeq
                G_F^2 ~~m_{\nu_\tau}^5 ~~V^2_{\tau e} / 192 \pi^3
  ~~~~~, }\eqno(10)$$
\no where $V_{\tau e}$ is a CKM-like
$\nu_e - \nu_\tau$ mixing and the above expression, valid for
$m_{\nu_\tau} >> m_e $, is slightly modified in the
$m_{\nu_\tau} \simeq  1 - 2$ MeV region
by a phase space factor. In general there could be
other non-standard model mechanisms contributing to
$\Gamma( \nu_\tau \rightarrow e^- e^+ \nu_e )$ so that, barring
accidental cancellation,
$\Gamma_{SM}^{( \nu_\tau \rightarrow \nu_e e^+ e^- )}$ is effectively a
lower bound to the rate of the process of interest. It has been pointed out
some time ago[13] that positron accumulation in the
Galaxy due to $\nu_\tau \rightarrow e^- e^+ \nu_e$ in all
previous
galactic supernovae
could lead to a strong 512 keV annihilation line. Demanding
that this will not exceed its observed value excluded
$\tau_{\nu_\tau} \geq 10^4 {\it sec}$ if
$\nu_\tau \rightarrow \nu_e e^+ e^-$ was the dominant decay mode.
More generally these results imply that
$${
{( { \tau_{\nu_\tau} \over { 10^4 {\it sec} } } )}^2
{}~~B( \nu_{\tau} \rightarrow e^+ e^- \nu_e ) \leq 1
  ~~~{\rm if} ~~\tau_{\nu_\tau}
\leq  2.7 \times 10^6 {\it sec} ~~~;}\eqno(11.a)
$$

$${
B( \nu_\tau
\rightarrow e^+ e^- \nu_e )
 \leq 10^{-5} ~~~~~~~{\rm otherwise} ~~~~~~~~,}\eqno(11.b)$$

\no with
$B( \nu_\tau \rightarrow e^+ e^- \nu_e )$
  the branching ratio for
the $e^+ e^-$ mode. In this paper, we consider the constraints
on the $\nu_\tau \rightarrow e^+ ~+~ e^- ~+~ \nu_e$ decay mode that arise
from the SN 1987A observations.

The observation of the neutrino pulses from Supernova 1987A (with a known
progenitor) has nicely confirmed the expectation for time scale, spectra
and overall energetics of the collapse and emitted neutrinos[14]. It puts
the above discussions on much firmer footing. The fact that
$\gamma$ ray fluence was measured
in an SMM satellite at the time of the neutrino burst
and shortly thereafter[15] has put stringent upper bounds[16] on the radiative
$\nu_\tau$ decays and also on the $e^+ e^-$ decay.

In section II, we consider in some detail
various photon generating mechanisms related to the
decay $\nu_\tau \rightarrow
 e^+ e^- \nu_e$ in the vicinity of a type II Supernova progenitor.
These could be used to improve the bounds in eqs.(11) found by considering
the comulative effects of past Supernovae.
Our discussion in section II
of photons emerging from annihilation of positrons and electrons
originating from $\nu_\tau$ decay parallels
  the analysis
of Dar and Dado[17] with which we are in basic agreement.
In particular,
the bounds are further improved in section II-c, in which annihilation of
positrons on ambient electrons are considered.
The analysis of II-b and II-c implies that
 even
if the $\nu_\tau$ decay occurs fairly near the star, there is
only a small probablity of $e^+, ~e^-$ annihilation.

  In section III, we speculate on the possibility that a $\nu_\tau$
excess, coupled with asymmetric $e^+, ~e^-$ spectra in the
$\nu_\tau \rightarrow
e^- e^+ \nu_e$ decay can lead to a large scale charge separation and
strong electric fields outside the progenitor.
This new ``Capacitor effect"
may have interesting observational implications, unless
$\Gamma( \nu_\tau \rightarrow e^+ e^- \nu_e )$ is strongly bounded.
Section IV contains the theoretical implications of our results.
In section V,
 we
  mention another possible effect of global neutrino spin allignment
around supernova.
We conclude in section VI.

\no {\bf {II. Constraints from $\gamma$-ray observation: }}~~~~~~

There are three mechanisms via which $\gamma$ rays can be
emitted in associated with $\nu_\tau \rightarrow e^- e^+ \nu_e$
decays near a type II supernova;
\item{a)} We have the radiative $\nu_\tau \rightarrow e^-~ e^+ ~\nu_e
{}~ \gamma$ decay;

\item{b)} Positrons and electrons emerging from the decays could
annihilate,

\no and finally

\item{c)} Positrons from $\nu_\tau$ decays
could annihilate on the
     ambient intersteller material near the
progenitor.

\no The lack
 of observation of such photons in the case of SN 1987A will limit
the space of the relevent parameters $m_{\nu_\tau} \equiv m, ~~ \tau_{\nu_\tau}
\equiv \tau$
and
$\Gamma( \nu_\tau \rightarrow e^- e^+ \nu_e ) \equiv \Gamma$ ( for
convienience we omit the suffixes ). The latter width can also be
expressed in terms of
the branching ratio B of the $e^- e^+$ mode
$${
\Gamma ( \nu_\tau \rightarrow e^- e^+ \nu_e ) \equiv \Gamma
= {B \over \tau } ~~~~, }
\eqno(12)$$
\no We will ${ \it a~ priori}$ limit the range of $m$ and
$\tau$ to

$${
 m \geq 2 ~~m_e \simeq 1 ~~{\rm MeV}~~~~~~~,}\eqno(13)$$
\no ( so that $e^+ e^-$ decay occures ) and
$$ {
\tau \geq 10 ~~{\it sec} ~~~~~~,}\eqno(14)$$
\no ( so that  $\nu_\tau$ can
decay outside the star ). The lifetime should not also be too long,
so that $\nu_\tau$ has time to decay before it arrives on the earth.

We point out here that cosmological arguments based on nucleosythesis
appears
to rule out $m_{\nu_\tau} \geq .3 ~{\rm to}~ .5$ MeV if
$\tau_{\nu_\tau} \geq 10^2 \sim 10^3 ~~{\it sec}$, provided
$\delta N_\nu \leq 0.6$[18] if one assumes the standard
big bang model of the universe.
 This overlaps with most of the range we are interested
in. If these arguments are taken literally, our bounds will be useful
only for
$10~~ {\it sec} \leq \tau_\tau \leq 10^2 \sim 10^3
 ~~{\it sec}$. We nonetheless
present our discussions as an independent piece of information
based on more direct
observations and no specific assumptions about the early universe.

During about 10 seconds when neutrinos are diffusing out of the
core ,
{\it i.e} during the time of the neutrino pulses,
we expect a total number of
tau neutrinos and antineutrinos $N_{\nu_\tau} \simeq 10^{58}$
with average energies ${\overline E_\nu} \simeq 20$ MeV, to be emitted.
The average $\nu_\tau$ decay will occur at laboratory lifetime $
{\overline \gamma}~
\tau$ after collapse and at a distance
  $${
R = \beta~ c~ {\overline \gamma} ~ \tau \simeq 6 \times 10^{11} {\tau \over m}
  ~{\it (c m)}~~~, }\eqno(15)$$
\no where ${\overline\gamma} = {\overline E}_\nu / m$
with $m$ in
MeV and $\tau, ~ \Gamma^{-1}$ in
second in all the subsequent,
 and
$\beta = {v \over c}$ with
$v$ being the neutrino velocity.

If at the time of decay, or shortly there after, a photon is
emitted towards the earth, it will be delayed relative
to the arrival of neutrino pulse by
$${
\delta t
 = (1- \beta ) {\overline\gamma}
 \tau \simeq {1\over {2 {\overline\gamma}^2} } {\overline\gamma} \tau
            = {\tau \over {2 {\overline\gamma}}} ~~~~.}
  \eqno(16)$$

Having set the general framework we now proceed to
discuss the specific photon generating
mechanisms (a - c) above.

\no  {\bf {II-a. Photons from
$\nu_\tau \rightarrow \nu_e e^+ e^- \gamma$
decay:}} ~~~~~

\no The rate of the radiative process
$\nu_\tau \rightarrow e^+ e^- \nu_e \gamma$ is expected to be that of
$\nu_\tau \rightarrow e^- e^+ \nu_e$ times
a factor ${\alpha \over {2 \pi} }  \simeq 10^{-3}$.
Since we have altogether ${ N_{\nu_\tau} } B~~~
e^+ e^-$ decays,  $10^{-3}{ N_{\nu_\tau} } B \simeq 10^{55}~B $ photons
should
arrive at earth during $\delta t \simeq {\tau \over {2 {\overline\gamma}} }$,
where $B$ is the branching ratio for
$\nu_\tau \rightarrow \nu_e ~e^+ ~ e^- $.
The local flux of photons would then be given by:
$${
\Phi_\gamma = { 10^{55}~ B \over { 4\pi d^2 ~~\delta t} }
\simeq {2\over {9 \pi}}~ 10^9 ~{\overline\gamma}~ \Gamma ~~~,}
\eqno(17)$$

\no where we use $d= 1.5 \times 10^{23} {\it cm}$
 for the distance from the earth to SN 1987A
and $\Gamma \equiv \Gamma( \nu_\tau \rightarrow \nu_e ~ e^+ ~e^- )
= B~ {\tau}^{-1}$.
Using ${\overline\gamma} = 20/ (m ~~{\rm in~~MeV})$, we have
$${
\Phi_{\gamma} \simeq {10^9 \over {(m~~{\rm in~~MeV})} }~~ ( \Gamma~~
{\rm in}~~{\it
sec}^{-1}) ~~{\it (cm)}^{-2}~{\it (sec)}^{-1}
{}~~~~~~~~~, }\eqno(18)$$
\no The SMM bound restriction[15] is $\Phi_{\gamma} \leq .1
{\it (cm)}^{-2}~ {\it (sec)}^{-1}$; hence it implies
$${
\Gamma \leq {( {10^{-10} \over m~~{\rm in~~MeV} } )}
 ~~{\it (sec)}^{-1} ~~~. }\eqno(19)$$

\no Note that the bound pertains directly to the rate of
$\nu_\tau \rightarrow e^+ e^- \nu_e$, as expected, since all other
decay modes do not contribute to the photons of interest.

\no {\bf {II-b. Photons from $e^+ e^-$
annihilation following $\nu_\tau \rightarrow \nu_e e^+ e^-$
decay: }}

\no In this subsection,
we discuss photons originating from mutual annihilation of
decay positrons and
electrons.
 The probablity that a $\nu_\tau$ will decay at a distance
between r and r + dr away from the star, $e^{- r /R} {dr \over R}$, is
roughly uniform
for $r < R$.
 The total number of electrons (positrons) from $\nu_\tau$
decays occuring inside a sphere r ( $r \leq R$ ) is therefore ( using
eq.15 ),
$${
{\cal N}^\pm (r) =  B~ { N} ( 1 - e^{-r/R} )
 \simeq {r \over R}~B~ 10^{58} = {B ~ r~ 10^{58} \over {\beta ~c~ \tau~
{\overline\gamma} } }
                    \simeq {\Gamma\over {\overline\gamma}} {10^{58}~r \over c}
{}~~~~, }\eqno(20)$$
\no where we have assumed $r << R$
and
the unit of $\Gamma ~~{\rm is}~~{\it (sec)}^{-1}$.

However, the original $\nu_\tau$ and descendant electrons move in the fixed
steller frame with a Lorentz factor $\overline\gamma$. Recalling that to a good
approximation, all $\nu_\tau$ are emitted at $t=0$, the
above ${\cal N}^\pm (r)$ electrons and positrons
will be within a shell of thickness
$\delta r$ given ( in analogy with eq.(16) ) by:

$${
\delta r \simeq {r \over { 2 {\overline\gamma}^2 } }~~~~~~.}\eqno(21)$$
\no The average number density in this shell is
$${
n^\pm (r) = { {\cal N}^\pm (r) \over {4 \pi r^2 \delta r} }
        \simeq { 10^{58}~~ {\overline\gamma}~ \Gamma \over { 2 \pi ~c ~r^2 } }
{}~~~~~.}\eqno(22)$$

\no The cross section for $e^+ e^-$ annihilation,
$e^+ e^- \rightarrow \gamma \gamma$, at center
mass energy $\sqrt s$ is related to
$\sigma_{Th} = 6 \times 10^{-25} {\it (cm) }^2$, the Thompson cross section
via:
$${
\sigma_{ann}( s ) = \sigma_{Th} { 4 m_e^2 \over s} = 6 \times 10^{-25}
                          { {( {\rm MeV} )}^2 \over s }
                           ~~{\it (c m)}^2 ~~~~. }\eqno(23)$$

\no For an electron and positron moving with energies $E^-, ~ E^+$, (which
for simplicity we take $E^{\pm} \geq m_e$ ) and at a relative angle
 $\theta_{1, 2}, ~~ s = 2 E^+ E^- ( 1- \cos\theta_{1, 2})$.
If ${\overline\gamma} >> 1$ all momenta tend to align in the
outward radial direction to within $\theta_i \simeq {1 / {\overline\gamma}}$.
Hence we can expand
$${
s \simeq E^+ E^- \theta_{1, 2}^2 \simeq { E^+ E^- \over {\overline\gamma}^2}
     \simeq {100 ~{\rm (MeV)}^2 \over {\overline\gamma}^2 }
{}~~~~,}\eqno(24)$$
\no and substituting in eq.(23) we finally have
$${
\sigma_{ann} \simeq 6 \times 10^{-27}~ {\overline\gamma}^2~~ {\it (c m)}^2
{}~~~~~~.}\eqno(25)$$

\no The probablity that any given positron among the
${\cal N}^{\pm}(r)$ in the shell will annihilate while traversing (in
essentially radial direction ) the shell of thickness $\delta r$
is
$${
P_{ann} ( \delta r ) \simeq \delta r~ n^{\pm}(r) ~\sigma_{ann}~
              {1\over { 2 {\overline \gamma}^2} }
{}~~~~~.}\eqno(26)$$

\no The last factor, ${( 2~ {\overline \gamma}^2 )}^{-1}$,
reflects the fact that the $e^+ e^-$ are
``chasing each other" on their joint outward drift. This reduces the relative
flux, in the ${\overline \gamma} >> 1$ limit, by
$1 / 2 ~{\overline \gamma}^2 $ which precisely compensates for the
increase of density due to the squeezing
from $r ~{\rm to}~ \delta r$ in eq.(22) by the same factor. Combining
eqs.(22) (25) and (26) we obtain
$${
P_{ann}(r) \simeq 0.8 \times 10^{20}~ { \Gamma \over { {\overline \gamma} ~r} }
  \simeq { 0.4 \times 10^{19}~ \Gamma \over r }
         ~~{( { m \over {\rm 1~MeV} })}~~~~~.}\eqno(27)$$

\no We are only interested in annihilations occuring outside
the progenitor,
\hfil\break
 ${\it i.e}~~  r \geq R_{st} \simeq 3 \times 10^{12}
                         {\it (c m)}$. Hence eq.(27)
  restricts  $P_{ann}$:
$${
P_{ann}(r) \leq 10^6~ \Gamma ~m ~~~~( {\rm all} ~ r \geq R_{st} )
 ~~~~~.}\eqno(28)$$

\no In particular, $P_{ann}(r) \leq 1$ if
  $${
    \Gamma \leq { 10^{-6} \over m}~~ { ( {\it sec} ) }^{-1}
{}~~~~.}    \eqno(29)$$

\no If eq.(27) yields an annihilation probablity
   $P_{ann}(r) \leq 1$ we can use this equation in computing
the total number of photons emerging from annihilations of
$e^+ e^-$ originating in sphere with radius r and thickness
$\delta r$
$${
{\cal N}_{\gamma}(r) = P_{ann}(r) ~{\cal N}^{\pm}(r)
         \simeq 2.5 \times 10^{67}~~ { \Gamma^2 \over {\overline \gamma}^2 }
{}~~~~~. }\eqno(30)$$

\no As expected ${\cal N}_{\gamma}$ scales with $\Gamma^2$. Its independence
of r
can be qualitatively understood as follows:
  if $r \rightarrow 2 ~r$ the
number of decays and the length for annihilation doubles but at the same
 time the original density of
electrons and positrons decreases as $1 / r^2$
 due to geometrical divergence.

The important point, in so far as putting stringent bounds on $\Gamma$ is
concerned, is that ${\cal N}_{\gamma} \simeq 2.5 \times 10^{67}~~ \Gamma^2 /
   {\overline \gamma}^2 $ photons are emitted within the
 times required for the
$\nu_\tau$'s to travel up to $2 R_{st}$ ( which we use as our running r value
),
and for the $e^+ e^-$ generated
in $\nu_\tau$-decay partially annihilate, yielding the above
${\cal N}_{\gamma}$ photons. These photons will arrive over a time spread
$${
\delta t  \sim 10 ~~{\it sec}
 ~~~~,}\eqno(31)$$
\no ( recall that there is a minimal 10 seconds width due to the
duration of the original $\nu$ burst. )

Arguing, as in the derivation of eq.(19) above, that the expected fluence
during 10 seconds after $t=0$ at earth
$${
\int_0^{\delta t } \Phi_{\gamma} dt \simeq { {\cal N}_{\gamma} \over
                { 4 \pi d^2 } } \simeq  10^{20}~~ {( { \Gamma \over {\overline
                       \gamma} })}^2 ~~~~~~, }\eqno(32)$$

\no should be smaller than the SMM bound of 0.6, we find
  $${
\Gamma \leq 0.8 \times 10^{-10}~ {\overline \gamma}
  = { 1.6 \times 10^{-9} \over {( m/ {\rm 1~MeV}) } } {( {\it sec} )}^{-1}
                                 ~~~~. } \eqno(33)$$

\no This last bound is the main result of this section. Since $\Gamma$
in eq.(33)
satisfies requirement for eq.(29) ( {\it i.e}
  $P_{ann} \leq 1$ ), our argument is self consistent.
 We can indeed show directly
that condition (29),
 ${\it i.e} ~~ \Gamma \leq {10^{-6} \over m} {( {\it sec
                   } )}^{-1}$,
should always be satisfied. If it is not satisfied there is a range of r
values
$${
R_{st} \leq r \leq r_{max} \equiv {( {\Gamma \over{ 10^{-6} / m}} )}
 ~~R_{st}~~~~~, } \eqno(34)$$
\no for which $P_{ann} = 1$. By considering the photons emerging from this
volume
( where all ${\cal N}^{\pm}(r_{max})$ annihilating to photon )
over a time interval $\Delta t = r_{max} / 2 {\overline \gamma}^2$,
we readily find out that the SMM bound is violated.

\no {\bf {II-c. Photons from
positron annihilation in
pre-supernova steller
debris:}}~~~

\no There is evidence[19] that the blue giant progenitor
sanduleak 6902
was emitting,
at the time of collapse, a wind of velocity
$v^B = 500 ~~km/ {\it sec}$ leading to mass loss at a rate
${\dot \mu}_B = 3. 10^{-6} M_{\odot} / yr$. The resulting density
near the steller surface is
$${
\rho_B |_{r = R_{st} } = { {\dot \mu}_{B} \over {
               4 \pi~ r_{st}^2 ~ v^B } } \simeq {3.~~ 10^{-12} g r \over
                     { (c m)}^3 } ~~~~. }\eqno(35)$$
\no Since
this density falls roughly like $1 / r^2 $
we expect an integrated column density of
$\int_{R_{st}}^{\infty} \rho ~ dr \simeq \rho_B |_{r = R_{st} }
                   R_{st} \simeq 1 {g ~ r\over {(c m)}^2 },
$ for
$R_{st} \simeq 3. 10^{12} c m$. [ Actually gravity diminishes somewhat
the velocity of escaping particles,
the density
$\rho(r)$ falls
slower than $1/ r^2$
and the column density
could be slightly higher
(by $\simeq 2$ ). ]
 This will introduce an uncertainty of almost a factor of two in our
estimates below.
$\nu_\tau$ decays occur between $R_{st}$
and $2~ R_{st}$ (say) at a rate
$${
{\dot{\cal N} }_{\nu_\tau \rightarrow
        e^+ e^- \nu_e}
\simeq 10^{58}~ {\Gamma \over {\overline\gamma}}
{}~e^{- { R_{st} \over { \beta~{\overline\gamma}~c~\tau} } }
{}~{( {\it sec} )}^{-1} ~~~~,} \eqno(36)$$
\no and this lasts for $\Delta t \simeq 10 \sim 100 ~{\it sec}$.
The fraction of positron annihilating in the $2~g r / {( {\it c m} )}^2$
(
or
$ 10^{24} / {( {\it c m} )}^2$ ) column is
$f = \sigma_{ann} \int n~ d r \simeq 0.1$ and
the annihilated photons would arrive within $100
 / 2~{\overline\gamma}^2
{\it sec}$ of the collapse (
for ${\overline\gamma} \leq 3$ only
otherwise
$\delta t \simeq 10$ {\it sec} ). Hence the
$\gamma$ ray flux on earth during this period is
$${
{
{\dot{\cal N}}_{ \nu_\tau \rightarrow e^+ e^- \nu_e}
{}~~f \over {4 \pi ~d^2~\delta t} } \leq 0.1 ~{\it sec}^{-1}~{\it cm}^{-2}
{}~~~~,} \eqno(37)$$

\no Taking $\delta t \simeq 10 ~{\it sec}$, then we have for the life time
satisfying ${\overline\gamma}~\tau \geq 100~ {\it sec}$

$${
\Gamma \leq {( { 10^{-10} \over {m ~~{\rm in~~ MeV} } })}
  ~~{\it sec}^{-1}~~~.} \eqno(38)$$
\no We note that, as we
proceed to further distance, the added column density and
annihilation probablity do fall off rather rapidly. The space
telescope discovered a ``ring" at a radius of
$R_{ring} \simeq 6 \times 10^{17} {\it c m}$
of thickness of
$\Delta R \simeq 10^{17}{\it c m}$
and particle number density of $
n = 2 \times 10^{4}/ {( {\it c m} )}^3$.
Presumably this ring results from the hot blue star wind
catching up with previous red giant ejecta. Note however that the extra
total column density in the ring
$\Delta R ~n \simeq 2 \times 10^{21}/ {( {\it c m})}^2
= 3 \times 10^{-3} g r/ {( {\it c m})}^2$
is negligible.

Thus even
if the $\nu_\tau$ decay occurs
very near to the steller surface most electrons and positrons
will escape.
In this connection we would like to comment on an early work of
R. Cowsik, D.N. Schramm and
P. H\"oflich[20] who
claimed a much stronger bound:
$${
\Gamma \leq 2.5 \times 10^{-16} ~~{ ({ \it sec})}^{-1}
{}~~( {\rm for} ~~m \leq k~T )~~~~~. } \eqno(39)$$
\no Their derivation assumes that the positrons and electrons lose all
their energy in the supernova ``debris".
However the main source of the ``debris"
is the steller envelope
ejected in the SN 1987A explosion.
The velocity of
 the envelope is low ( $\simeq 3000 ~km/
{\it sec} = 10^{-2}~c$ ). It will
therefore never catch up with the $\nu_\tau$
and/or their $e^+ ~e^-$
decay products which are moving effectively with the speed of light.
Cowsik et al[20] suggest that
there could be trapping of the decay electrons and positrons in the galactic
magnetic fields of order $5~\mu_{G}$. We fully agree that there will
be
magnetic
trapping on a galactic scale. There cannot,
however,
even for $\Gamma$ is as small as $10^{-11}$, be a local trapping of the
$e^+ ~e^-$ in the neighbourhood of the supernova so that
the ``debris" could eventually catch up with them, allowing the
stringent bound to be reinstated.
To see it, let us assume that:
$${
\Gamma \simeq 10^{-11} ~~~~~~~. } \eqno(40)$$
\no The total number of
$\nu_\tau$'s decaying to
$e^+ e^-$ within a year would be
$10^{58} \cdot 10^{-11} \cdot 3 \times 10^7 \simeq 3 \times 10^{54}$
and the total $e^+ e^-$
energy would be $\simeq 2 \times 10^{54} \cdot 20 ~{\rm MeV}~
= 6 \times 10^{49}$ erg.
Because of their large energy and momentum, the flux of positrons
will simply blow the magnetic field away. Indeed
if the distance scale over which trapping occurs is denoted by
$R^*$, then
$R^*$ can be estimated by
requiring that the total B field energy swept satisfies:
$${\int E \cdot dV = ~{ 4 \pi ~R^{*^3} \over 3}~ {B^2 \over {8 \pi} }
       ~ \simeq {1\over 6} R^{*^3} ~ 10^{-12}
           = 6 \times 10^{49}{\rm ergs} ~~~~,
} \eqno(41)$$

\no leading to $R^* \simeq 10^{21} {\it c m}$ ( while the nearby
steller field is intense, say $B_{steller} \simeq 10^3$ Gauss, its total
energy
${4 \pi \over 3} ~R_{st}^3 ~{ B^2\over {8 \pi} } = 10^{44}$ is smaller ).
Thus, in our opinion the stringent bounds suggested in ref.20 is unlikely
to hold.

\no {\bf {II-d.
$\nu_\tau$ decays inside the steller envelope:}
}

\no In this subsection we consider the case where $\nu_\tau$ decays
inside the star. If most of the decays occure inside the
star, the bound in eq.(38) will be weaken:
$${
\Gamma  \leq {( { 10^{-10} \over { m~~ {\rm in~~MeV} }})}
 ~~e^{ 100
        \over { {\overline\gamma} ~\tau } }
{}~~{( {\it sec} )}^{-1}
{}~~~~.  } \eqno(42)$$
\no But, if $\tau = 10~ {\it sec}~~
{\rm and}~~{\overline\gamma} =2$,
we still have an appreciable bound
$\Gamma \leq 10^{-9} $.

It is amusing to note that if the decay distance
$R \simeq {\overline\gamma}~c~\tau$ falls somewhat near the surface
$R=R_{st}$ there could be yet another independent effect and certain ranges of
$\Gamma$ could be excluded even without apealing to the SMM results.
The point is that all models
of the progenitor
(see {\it e.g} Barkat and Wheeler[21]) suggest that
near the steller surface say between $R_{st}/2$ and
$R_{st}$ there is, thanks to
a much reduced intensity, a relatively small mass [ $.5 ~~M_{\odot}$ ].

The total energy deposited into the layer is
\hfil\break
 $
W_{dep}
\simeq 10^{53} ~{\rm erg}~
B ~ {R_{st} \over {2~ {\overline\gamma} ~c ~\tau} }
{}~e^{ - { R_{st} \over {2~{\overline\gamma}~c~\tau} } }$.
It should be compared to the gravitational binding of the layer
$${
B.E. =
- G_N{ M_{st} ~\Delta M \over {R_{st} ~3/4} }
\simeq 4 \times 10^{47} ~~{\rm erg} ~~~~, }  \eqno(43)$$
\no where we use $M_{st} \simeq 10 M_{\odot}~~{\rm and}~~
R_{st} = 3 \times 10^{12} {\it cm}$. \hfil\break
 If $W_{dep} \geq B.E. ~~{\it i.e.}~~
{\rm if}~~ B ~ x ~e^{-x} \geq 4 ~10^{-6}
{}~~{\rm with}~~ x = R_{st} /2~{\overline\gamma}~c~\tau$,
the whole layer would be blown off.
 Since this process would start right away
after the $\nu$ pulse, we should have spotted dramatic changes in
the steller luminosity during period between the $\nu$ pulses and the
supernovae explosion. We therefore conclude that $B x e^{-x}
\leq 4 \times 10^{-6}$, which gives an upper limit
$${
\Gamma \leq { 10^{-6} \over {( m ~~{\rm in~~MeV} )} }
{}~~~~~. } \eqno(44)$$
\no This is a much weaker constraint on $\Gamma$ than what we got before.

\no {\bf {III. Capacitor effect induced by
$\nu_\tau \rightarrow \nu_e e^+ e^-$ :}}

\no It is amusing to note that the
$\nu_\tau \rightarrow e^- e^+ \nu_e$ decay of the many $\nu_\tau$'s
produced in a type II supernova, coupled with a plausible $\nu_\tau -
{\overline
\nu_\tau}$ asymmetry could lead to a rather fascinating possibility.
Specifically, we could have a large scale coherent phenomenon of charge
separation in the neighborhood of the star and a corresponding
build-up of large scale strong radial electric fields. We will investigate
this phenomenon in more detail in the future, but would like to
briefly mention the essential features here.
There are three main ingredients which conspire to
make such an effect conceivable:
\item{i)} The likely $\nu_\tau - {\overline{\nu_\tau}}$ asymmetric emission;

\item{ii)} The different energy spectra of electrons and positron emerging
from a decay of tau neutrinos due to the V-A nature
of the interaction;

\item{iii)} The likely survival of most electrons and positrons from $\nu_\tau$
and
$\overline{\nu_\tau}$ decays
in the vicinity of the progenitor star.

\no Let us elaborate on these points:

\no i) ~~~~~~The collapsing stellar core of mass
$\simeq 2~ M_{\odot}$
has a large electron (lepton) number
$${
N_L \simeq N_e \simeq 10^{57} ~~~. } \eqno(45)$$

\no The collapse is facilitated by a copious production
of electron neutrinos via the electron capture (weak) process:

$${
e^- ~~p \rightarrow n ~ \nu_e
{}~~~, } \eqno(46)$$

\no which lowers the Fermi energy of the electrons. Some portion of these
neutrinos are emitted right away during the few mili-seconds
of the collapse as a ``neutronization burst"
 preceeding the main $\sim 10 ~{\it sec}$
emission of the thermal neutrinos. However as the star quickly collapses
and core densities $\rho
\geq 10^{11} g r/ {(c m)}^3$ are achived, the
electron neutrinos become trapped and we have thermal equiliblium of
reaction in eq.(46) and its inverse
$n ~ \nu_e \rightarrow e^{-} ~~ p $. If we have
processes which conserve overall lepton number but allow the interconversion
of different leptonic flavors, then the free energy of the system could be
further lowered by having the trapped lepton excess reside in $\nu_\mu ,~~
\nu_\tau$ as well.

Thus for massive mixing neutrinos we could have along with reaction in eq.(46)
also:
$${
e^- ~~p \rightarrow \nu_\tau ~~n
{}~~~, ~~( {\rm or}~~ e^- ~~p \rightarrow \nu_\mu ~~n ) ~~,
}\eqno(47)$$

\no with cross sections
reduced by the appropriate $\theta_{ e \tau}^2$
( or
$\theta_{e \mu}^2$ ) factors due to CKM-like mixing angles. While
$\theta_{e \mu}$ and
$\theta_{e \tau}$ are likely to be small, there are many weak interaction
reactions of the type $e^- ~~p \rightarrow n~~ \nu$. Indeed for density
$\rho \simeq 10^{15} g r / {(c m)}^3$ and energy
$E \geq 30$ MeV involved, the mean free path
 for weak reactions
is very short $l \simeq 30 ~{\it c m}$. As neutrinos
diffuse out of the core of radius
$r_c \simeq 10 ~km$ they suffer many collisions
$${
{ N}_c \simeq {( {r_c \over l} )}^{2} \simeq 10^{9}
{}~~~. ~~} \eqno(48)$$
\no This is also the relevent number of $e^- ~p
\rightarrow n ~ \nu$ reactions. Thus we expect an initial electron flavor
excess ${ N}_e$ to build up
gradually a $\nu_\tau , ~~(\nu_\mu
)$ excess of magnitude
$${
\Delta { N}_{\nu_\tau}
 = { N}_{\nu_\tau} - { N}_{\overline{\nu_\tau}}
\simeq { N}_c ~~ \theta_{e \tau}^2 ~~ { N}_e
\simeq ( { N}_c ~ \theta_{e \tau}^2 ) ~~10^{57}
{}~~~~, } \eqno(49)$$

\no ( and $ \Delta { N}_{\nu_\mu} \simeq {
N}_c ~ \theta_{e \mu}^2 ~~ 10^{57}$ ).

Obviously these expression hold only if ${ N}_c ~\theta^2 < 1$, otherwise
we would have equilibration of the different flavor excess
$${
\Delta { N}_{\nu_\tau} = \Delta { N}_{\nu_\mu}
=....= {1\over 3} { N}_e \simeq 0.3 ~~10^{57}
{}~~~~, } \eqno(50)$$

\no Defining in general $\epsilon = ({ N}_{\nu_\tau}- { N}_{{\overline
\nu}_{\tau}}) / ({ N}_{\nu_\tau} + { N}_{{\overline\nu}_{\tau}})$,
we see that we have
appreciable asymmetry
$\epsilon \geq 10^{-3}$ even if
$\theta_{e \tau}^2 \simeq 10^{-12}$.

Evidently, new physics, flavor violating processes such as \hfil\break
$\nu_\mu ~+~ \nu_\mu
\rightarrow \nu_\mu ~+~ \nu_\tau$,
if present, could even enhance the $\nu_\tau$ flavor asymmetry via
$\nu_e \rightarrow \nu_\mu \rightarrow \nu_\tau$.
In any case, a build up of a
$\nu_\tau$ excess
from an initial $\nu_e$ excess
is quite plausible.
  Let us study the consequence of this
$\nu_\tau$ excess, if the decay
$\nu_\tau \rightarrow \nu_e e^+ e^-$ occur with a signaficant
branching ratio.

ii) ~~~~~~The decay $\nu_\tau \rightarrow
        e^- e^+ \nu_e$
is mostly likely caused by
 V-A interactions (such as W-exchange or
$\Delta^0$ Higgs exchange), and leads to momenta of final
$e^-$, which are approximately twice of that of $e^+$.
This asymmetry is preserved under boosts of the
original $\nu_\tau$
from rest to $E_{\nu_\tau} \simeq 20 $ MeV (say).
Only if
we have equal number of $\nu_\tau$ and
${\overline{\nu_\tau}}$ and with equal energy spectra,
{\it e.g} in the ideal thermal case,
 could we argue, on general (CP) symmetry grouds, that the generated
$e^+$ and
$e^-$ population will have the same features.
Otherwise, we will in general have $e^+$ and
$e^-$
from the $\nu_\tau$ decay moving with different momenta.

On average an $e^- ~( e^+ )$ emerging from a
$\nu_\tau$ decay will have a velocity
$${
{\overline{\beta^{\pm}}} = 1 - {1 \over {2 ~ {\overline{\gamma_{\pm}^2}}}}
{}~~~~~~, } \eqno(51)$$

\no with ${\overline{\gamma_{\pm}}} \equiv {\overline{E}_{\pm}}/m$ having
an overall average value ${\overline\gamma}=
         { {\overline\gamma}_+ + {\overline\gamma}_- \over 2} \simeq
              { {\overline{E_+}} + {\overline{E_-}} \over 2}
         \simeq { 7 ~{\rm MeV} \over {1/2 ~{\rm MeV}} }
            = 14 $. (Recall that
$E_{\nu_\tau} = 20$ MeV is
the initial energy,
 shared by the three light leptons, and
averaging over the asymmetry yields then ${\overline{E}} \simeq {E_{\nu_\tau}
/3 } \simeq 7$ MeV.)
The energy-momentum asymmetry implies that the $e^+~ e^-$
originating from $\nu_\tau$'s have on average a velocity difference
$${
\Delta \beta = \beta_{e^-} - \beta_{e^+} \simeq
           {1\over 2} {1\over {2 ~ {\overline\gamma^2} }}
\simeq 2 \times  10^{-3} ~~~. ~} \eqno(52)$$

iii) ~~~  The discussion in section
 II-b and II-c  implies that the probablity of
positron annihilation is smaller than $\simeq 10^{-1}$.

Let us proceed to show then how the ingredients i) and
ii) above can be synthesized to
make a scenario for large scale charge separation and
corresponding strong
electric fields. The basic point is simple. There is a systematic tendency
of the
 $\Delta N_- = \epsilon~ B ~N_{\nu_\tau}$
 electrons coming from the decay of the
excess ``$\nu_{\tau}$"
to move radially out with a velocity exceeding that of the corresponding
 positrons by $\Delta \beta \simeq 2 \times 10^{-3}$.
Let us make the crude approximation that all electrons and positrons emerged
from $\nu_\tau$ decay
at $ t = \beta~ {\overline\gamma}~ \tau$ and
$r = R = c t$. Further, let us ignore, in the first approximation, the response
of the (almost) neutral plasma of $n_{e^+} + n_{e^-}$. After time
$\Delta t$ we have the two layers of charge
$\pm Q,~~ ( Q = \Delta N^{- (+)} ~ e )$, separated by a distance $\Delta r
= \Delta\beta ~ c ~ \Delta t$.
Even for $\Delta N^{- (+)} << N_{\nu_\tau}$
{\it i.e} for
$ \epsilon ~B ~<<~ 1$ such putative separation of charges
will generate large electrostatic fields and energies. This is so
because the magnitude of the separating charges
  $${
Q = \epsilon ~ B~ 10^{58} ~e~ ~ = {\epsilon ~~B \over 2} 10^{49}
   ~~{\it esu} ~~~, } \eqno(53)$$
\no is very large. There will therefore
be strong responses (in particular of the
neutral $e^- e^+$ plasma ) which will tends to quench these
fields and the charge separation. However in the process the
``asymmetry" energy ( {\it i.e} the energy of the collective relative
$e^-, e^+$ motion )
$W_{A}
\simeq \Delta N_- \times 7 ~{\rm MeV} \simeq \epsilon ~ B ~ 10^{53} {\it erg}$
would be dissipates via some e.m., fluorercence effect and could allow for
yet another sensitive signature of
the
$\nu_\tau \rightarrow e^- e^+ \nu_e$ decays.
We do not derive any specific upper bound on $\Gamma$ from
these considerations, due to a lack at present of a quantitative study of
the charge seperation.

\b
\no {\bf {IV. Theoretical
implication of the bound on
$\Gamma ( \nu_\tau \rightarrow \nu_e e^+ e^- )$ :}
}~~~

\no  Let us now discuss the theoretical implications of the
stringent bounds on the decay rate for $\nu_\tau \rightarrow
\nu_e ~ e^+ ~ e^-$. This decay can occur if
 the standard model is extended by the inclusion of three
right-handed neutrinos (one per family), so that the neutrinos have masses
and
there will be mixings between different generations.
The decay rate is given in eq.(10).
The upper bound on $\Gamma$ in eq.19 implies that
$${
V_{\tau e} \leq 4 \times 10^{-3} {( { {\rm MeV} \over m_{\nu_\tau} })}^{3}
{}~~~~. }\eqno(54)$$

\no In general, there must be mechanism for an invisible decay $\nu_\tau$
with lifetime in the range 10 {\it sec} to
$10^8$ {\it sec} ( so that
$\nu_\tau$ decays outside the supernova core and before reaching the earth
). This can be the case in the minimal singlet Majoron model[7].

It must also be noted that, a Majorana $\nu_\tau$ will contribute to
neutrinoless
double beta decay via the mixing $V_{\tau e}$. For
$m_{\nu_\tau}$ in the few MeV range, we have the bound[22]
$V_{\tau e}^2 ~ m_{\nu_\tau} \leq 1 ~{\rm eV}$. This bound is consistent
with eq.54 for $m_{\nu_\tau} >$ 2 MeV.

 A specific model where neutrinos are massive
is the left-right symmetric model with the see-saw mechanism for
neutrino masses[1]. For a $W_R$-mass in the TeV range,
this theory predicts a tau
neutrino mass in the MeV range
as already mentioned. Therefore, the decay
$\nu_\tau \rightarrow \nu_e ~ e^+ ~ e^-$ is kinematically allowed in this
model.
In addition to the already mentioned
$W_L$-exchange contribution that arises from
$\nu_e - \nu_\tau$ mixing, there are also Higgs contributions
to the process $\nu_\tau \rightarrow \nu_e ~e^+ ~ e^-$ in this model.
Furthermore, if certain Yukawa couplings are chosen to be of order
$\leq 10^{-1}, ~~\nu_\tau$ will have a lifetime longer than 10 {\it sec},
so that, we can apply the discussions of the previous sections.
The SN 1987A upper bound on this process therefore leads to upper bounds on
these Higgs couplings, which in turn imply upper bounds on the rare process
$B( \tau \rightarrow 3 ~e )$ for
$m_{\nu_\tau} \geq 2$ MeV since the same
couplings are involved in the $\tau \rightarrow
3~ e$ decay.
To discuss this, let us write down the coupling of the left-handed triplet
Higgs in the basis where all leptons are mass eigenstates:
$${
\eqalign{
{\cal L}_Y=& \nu_L^T K F K^T C^{-1}
           \nu_L \Delta^0 - {\sqrt 2}
            \nu_L^T K F C^{-1} E_L \Delta^+ \cr
            & - E_L^T C^{-1} F E_L \Delta^{++} ~~+ ~{\it h.c} \cr}
       ~~~,}  \eqno(55)$$
\no where $\nu = ~( \nu_e , ~ \nu_\mu , ~ \nu_\tau );
          E = ( e, \mu , \tau )$
and F is the a real symmetric $3 \times 3$ matrix and K is the neutrino mixing
matrix. In a low scale
$M_{W_R}$
theory (
$M_{W_R} \simeq {\rm TeV}$ ), the masses of the left-handed
triplets $\Delta^0_L, ~~\Delta^+_L ~~{\rm and}~~\Delta_L^{++}$ can be in the
100
GeV range without unnatural fine tuning of the scalar self couplings. We set
the $F_{e \mu}=0$ in order to prevent
$ \mu \rightarrow 3 e$ decay, whose branching ratio has an experimental upper
limit of $10^{-12}$[23]. The present experimental upper limit on
$\mu \rightarrow e \gamma$ also imposes strong constraints on the
product
$ F_{e \tau} ~F_{\mu \tau} \leq 2 \times 10^{-5}$. The process
$\nu_\tau \rightarrow \nu_e ~e^+ ~ e^-$ arises in this model at the tree
level from $\Delta_L^+$ exchange
and leads to:
$${
\Gamma( \nu_\tau \rightarrow \nu_e ~ e^+ ~ e^- ) \simeq
 {  {( F_{e \tau} + K_{1 3}F_{\tau \tau} )}^2 ~F_{e e}^2 ~ m_{\nu_\tau}^5
\over {1536 \pi^3 ~M_{\Delta_L^+}^4 }} ~~~~.} \eqno(56)$$

\no Using the bound in eq.19, we then get
$${
{( F_{e \tau} + K_{1 3} F_{\tau \tau} )}^2 ~F_{e e}^2 \leq
                  4.6 \times 10^{-14}~~{( { {\rm MeV} \over
                    m_{\nu_\tau} } )}^6 ~~{( {M_{\Delta^+_L}
               \over {\rm GeV} })}^4~~~
                       .} \eqno(57)$$

\no Barring unnatural cancellation between $F_{e \tau}~~{\rm and}~~
              K_{1 3} F_{\tau \tau}$, we conclude that,
  $${
F_{e \tau}^2 ~ F_{e e}^2 \leq 4.6 \times 10^{-14}
                    {( { M_{\Delta_L^+}\over {\rm GeV} })}^2 ~
 {( { {\rm MeV} \over m_{\nu_\tau} })}^6
                ~~~~~.  }  \eqno(58)$$

\no A question that now arises is whether this constraint suppresses the
$\nu_\tau$ decay rate so much that an MeV range
$\nu_\tau$ fails to satisfy the mass density constraints.

Let us therefore study the total decay rate
for $\nu_\tau$. In a general left-right models,
eq.(54) implies that, \hfil\break
$\nu_\tau \rightarrow {\overline \nu_\mu} \nu_e \nu_e , ~~
      {\overline \nu_\mu} \nu_\mu \nu_\mu , ~~ {\overline \nu_e}
                     \nu_\mu  \nu_\mu , ~~ {\overline \nu_e}
                       \nu_e \nu_e $ amplitudes are governed by the
following combination of couplings:
$${
{\overline \nu_\mu} \nu_e \nu_e : ~~~~~~
\simeq F_{e e}~ ( F_{ \tau \mu} + K_{3 2}F_{\mu \mu} + K_{2 3}F_{\tau \tau}
                            ) ~~~~; }\eqno(59.a)$$

$${
{\overline \nu_\mu} \nu_\mu \nu_\mu : ~~~~~~
        \simeq F_{\mu \mu} ( F_{\tau \mu} + K_{3 2}F_{\mu \mu}
                      + K_{2 3} F_{\tau \tau}) ~~~~; } \eqno(59.b)$$

$${
{\overline \nu_e} \nu_\mu \nu_\mu : ~~~~~~
           \simeq F_{ \mu \mu} ( F_{ \tau e} + K_{3 1} F_{e e} + K_{1 3}
              F_{\tau \tau}
                ) ~~~~; }  \eqno(59.c)$$

$${
{\overline \nu_e} \nu_e \nu_e : ~~~~~~
\simeq F_{e e} ( F_{\tau e} + K_{3 1} F_{e e}
                         + K_{1 3} F_{\tau \tau}
                       ) ~~~~. } \eqno(59.d)$$

\no Note that if we assume the diagonal F-couplings
$F_{e e}, ~F_{\mu \mu}, ~ F_{\tau \tau}$ are unsuppressed,
supernova constraints require only
$F_{\tau e} + K_{1 3}F_{\tau \tau}$ to be highly suppressed; however, dominant
contributions to $\nu_\tau \rightarrow 3 ~ \nu$ decay can still arise
from
$F_{\tau \mu}, ~~K_{3 1}F_{e e}, ~~K_{3 2}F_{\mu \mu}$ terms in the above
equation to make the MeV $\nu_\tau$ life time
consistent with cosmological mass density constraints. For instance, we require
$${
F_{a a}~F^{\prime}_{b c} \geq
5.7 \times 10^{-9} {( { 10 ~{\rm MeV}\over m_{\nu_\tau} })}^{3/2}
                ~~{( {M_{\Delta_L^0} \over {\rm GeV} })}^2
{}~~~~~~,   } \eqno(60)$$
\no where $a = e, \mu ~~{\rm and} ~~F^{\prime}_{b c}$ is
either
$F_{\tau \mu}, ~~K_{3 1}F_{e e} ~~{\rm and}~~K_{3 2}F_{\mu \mu}$.
These lower bounds are consistent with the upper bounds on the
neutrino mixing angles.

Let us now discuss some other implications. We note that the combination
of couplings $F_{e \tau}F_{e e}$ also leads to the rare decay
of tau-lepton, $\tau \rightarrow 3~ e$, via
$\Delta^{++}_L$ exchange. We can therefore derive the following
inequality:
$${
B( \tau \rightarrow 3 ~e ) \leq 7 \times 10^{-6}
               {( { {\rm MeV} \over m_{\nu_\tau} })}^6 ~
                     {( { M_{\Delta_L^+} \over M_{\Delta_L^{++}} })}^4
{}~~~~~.
} \eqno(61) $$
\no In the left-right model, we have
$${
{( { M_{\Delta_L^+}^2 \over M_{\Delta_L^{++}}^2 })}^2 ~ = ~ { 1 + \alpha
                            \over {1 + 2 \alpha} }
                           ~ \leq 1 ~~~~~, } \eqno(62)$$
\no where $\alpha$ is a real positive parameter. Using this, we
get an upper bound on $B( \tau \rightarrow 3~ e )
\leq 7 \times 10^{-6}
{( { {\rm MeV} / m_{\nu_\tau} } )}^6$
for
$m_{\nu_\tau} \geq $ MeV.
In order to see the significance of this bound, we observe that, if the upper
limit of the branching ratio $B( \tau \rightarrow 3~ e )$
keeps going down, no conclusion can be derived from it; on the other hand, if
evidence for
$\tau \rightarrow 3 ~e$ is found, then, it would rule out
an MeV range $\nu_\tau$ in the
framework of the minimal left-right symmetric model.

\no {\bf {V. Possible neutrino spin alignment during
neutronization:}}

\no In this section, we speculate on
 yet another possibility where the unique
parity violating character of the weak interactions conspires to make
macroscopic spin allignment of neutrinos.
For this purpose,
 let us consider the initial ``neutronization" pulse of
electron neutrinos and focus our attention on some region
in the collapsing core corresponding to a small
solid angle and bounded between two radial shells $R_1 \leq r \leq R_2$.
Let us assume that most of the neutronization neutrino traversing this region
have originated in a smaller core region yet, $r << R_1$.
The trajectories of the neutrinos will tend therefore to predominatly
align in the (outward) radial direction. Because, all these are
left-handed  neutrinos
(rather than an equal mixture of neutrinos and anti-neutrinos)
of negtive helicity,  the spins of these neutrinos will align too. Thus if
we have
at a given time
$N_{\nu}$ neutrinos in this region,
 the total spin of the neutrinos will be
cohenrently added to a large, macroscopic, total angular momentum:
$${
S_{\nu} = N_{\nu} ~{\hbar \over 2}
{}~~~~~. }  \eqno(63)$$

\no The corresponding densities
$s_\nu,  ~n_\nu$ are obtained by dividing by
\hfil \break
 $( R_2 - R_1 ) R_1^2 d \Omega
\simeq V$ the volume of the region in question. Assume that the neutronization
pulse containing altogether
$\sim 0.5 \times 10^{57}$ neutrinos lasts
for a milisecond and originates from a region of corresponding size
say $= c \delta t \simeq 10^8 {\it cm}$ then
$${
n_\nu \simeq { 5 \times 10^{56} \over {4 \pi {[ 10^8 ]}^3 }}
\simeq 4 \times 10^{31} ~/{( {\it cm} )}^3 ~~~~. } \eqno(64)$$
\no Let us consider possible implications of such a
  macroscopic spin
allignment. Two
possibilities come to mind: one,
if the neutrino has an intrinsic magnetic dipole
$\mu_{\nu_e}$, then, naively, we would have expected
a build-up of a local ``neutrino magnetization" magnetic field of magnitude
$B = 4 \pi n \mu_\nu$. This however is not the case. The would be
field is radial and Maxwell equation $div {\vec B} = 0$ excludes such
fields implying a complete cancellation
of the spherically symmetric field\footnote{[F.2]}{
In principle the $\nu_e$ could have also an electric dipole moment and
 radial electric fields are allowed. However for this to be the case
it needs also CP violation in the neutrino sector.}.

Another interesting possibility concerns the singlet Majoron model.
In this model we have a massless boson
namely the Majoron $\chi$. The Majoron couplings to ordinary
light neutrinos are small:
$${
{\cal L} \simeq {m_i \over V_{BL} }
               \chi {\overline\nu} \gamma_5 \nu
{}~~~,  }  \eqno(65)$$
\no where $m_i$ the neutrino mass and
$V_{BL}$ the scale of B-L symmetry.
For $\nu_e, ~~m_{\nu_e} \leq 10 {\rm eV}$
and $V_{BL} \geq 10^2 \sim 10^3$ GeV, we have ${m_i \over V_{BL} }
\leq 10^{-10} \sim 10^{-11}$, which is very small. Furthermore the
latter coupling involves, in the non-relativistic limit
of soft Majoron emission (from neutrinos), momentum dependent spin flip
interaction ( and correspondingly generates a
$1/ r^3$ spin-dependent
$\nu \nu$ potential ).
Hence such interactions are almost virtually undetectable.
The situation might be more favorable in the case when all neutrino spins are
alligned. Even in this case the $1/ r^3$ fall off makes the
effect of Majoron mediated $\nu_e ~ \nu_e$ interaction negligible.

\b
\no {\bf {VI. Conclusion:}}

\no In this paper, we have studied the possible effects connected with
$\nu_\tau$'s emitted in
SN 1987A if their masses are in the MeV range and the
electroweak theory allowes the visible decay
$\nu_\tau \rightarrow \nu_e e^+ e^-$. We derive constraints on this decay mode
of $\nu_\tau$, if the
$\nu_\tau$-lifetime is longer than 10 {\it sec}. As mentioned in the text,
 our results agree (where they overlap) with the earlier work of Dar and
Dado[17].
 We explain the physics underlying the derivation of these constraints.
We study the implications of these constraints on the Higgs couplings of the
left-right symmetric models with a low scale see-saw mechanism.
We then
speculate on two new effects, which could be generated during the early
``neutronization" moment of the supernovae, if there is non-negligible
$\nu_e - \nu_\tau$ mixing: one of them has to do with the formation
of a ``giant capacitor" in intergalactic space and another with a
possible large
neutrino spin alignment near the supernova core.

\bbb
\bb
\b
One of us (S.N.) thanks the nuclear theory group at the University of
Maryland for hospitality.

\bbb
\bb
\filbreak
\bb
\b

\b

\ce{\bf References}
\b
\item{[1]}
For a review of the situation, see R.N. Mohapatra and
P.B. Pal, ``{\it Massive Neutrinos in Physics and
Astrophysics}", world scientific (1991).

\item{[2]}M. Gell-Mann, R. Ramond, R. Slansky, in {\it Supergravity},
eds. D. Freedman and
P. Van Nieuenhuizen, Amsterdam, North Holland (1979);
T. Yanagida, in {\it
Proceedings of Workshop on the Unified theory and
the Baryon number of the Universe}, (KEK, Japan, Febuary 1979) ed.
O. Sawada et al; R.N. Mohapatra and
G. Senjanovi\'c, Phys. Rev. Lett. {\bf 44},
912 (1980).

\item{[3]}R.N. Mohapatra and G. Senjanovi\'c , Phys. Rev. Lett.
 $\bf 44$, 912 (1980); Phys. Rev. $\bf D23$, 165 (1981).

\item{[4]}
    D. Kreinick, in {\it Beyond
the Standard Model III}, Ottawa, Canada, June 22-24, (1992).

\item{[5]}D. Dicus, E. Kolb and V. Teplitz, Phys. Rev. Lett.
   $\bf 39$, 168 (1977).

\item{[6]} G. Steigman and M. Turner, Nucl. Phys. $\bf B253$, 375 (1985).

\item{[7]}Y. Chikashige, R.N. Mohapatra and R.D. Peccei,
Phys. Lett. {\bf 98B}, 265 (1981).

\item{[8]}J. Schecter and J.W.F. Valle, Phys. Rev. {\bf D25}, 774 (1982).

\item{[9]}
P. Herczeg and R.N. Mohapatra, Phys. Rev. Lett. $\bf 69$, 2475 (1992).

\item{[10]}R.N. Mohapatra, S. Nussinov and X. Zhang, UMDHEP 94-30,
         August 1993.

\item{[11]}D. Krakauer et al, Phys. Rev. {\bf D44}, R6 (1991).

\item{[12]}For a review and references, see G. Raffelt, Phys. Rep.
                    {\bf 198}, 1 (1990).

\item{[13]}A. Dar, J. Goodman and S. Nussinov, Phys. Rev. Lett.
            {\bf 58}, 2146 (1987).

\item{[14]}A. Burrows and J. Lattimer, Ap. J. {\bf 307}, 178 (1986);
           R. Mayle, J. Wilson and D. Schramm, Ap. J. {\bf 318},
   288 (1987).

\item{[15]}D. Chupp, C. Reppin and W. Vestrand, Phys. Rev. Lett. {\bf 62},
                       505 (1989).

\item{[16]}L. Oberauer and F. von Feilitzsch, Rep. Prog. Phys.
{\bf 55}, 1093 (1992);

\item{[17]}A. Dar and S. Dado, Phys. Rev. Lett. {\bf 59}, 2368 (1987).

\item{[18]}E. Kolb et al, Phys. Rev. Lett.
                    {\bf 67}, 533 (1991);
 I. Rothstein and A. Dolgov, Phys. Rev. Lett. {\bf 71}, 476 (1993).

\item{[19]} {\it {``Supernovae" Tenth Santa Cruz Summer Workshop
             on Astronomy and Astrophysics}} ed. S.E. Woosley;
            { \it {ESO/EPC Workshop on SN 1987 A and
                      Other Supernovae}} Isola D'ella (1990),
ed. I.J. Danziger and I. Kodsai.

\item{[20]}R. Cowsik, D.N. Schramn and P. H\"oflich, Phys. Lett. {\bf 218B},
             91 (1989).

\item{[21]}Z. Barkat and J.C. Wheeler, Ap. J.,
             {\bf 332}, 247 (1988).

\item{[22]} For a recent review, see H. Klapdor, Heidelberg Preprint
          MPI-HV-34-92 (1992).

\item{[23]}Review of Particles Properties, Phys. Rev. {\bf D45}, no.11,
              (1992).

\bye